\pdfoutput=1
\documentclass[10pt,onecolumn]{IEEEtran}
\usepackage {amssymb,rotating,graphicx,cite, calc, color, psfrag}
\usepackage{authblk}
 \usepackage{booktabs}
 \usepackage[table,xcdraw]{xcolor}
\usepackage[ruled,norelsize]{algorithm2e}
\usepackage{balance}

\usepackage[cmex10]{amsmath}
\usepackage{amssymb,amsthm,amsfonts,mathrsfs}
\usepackage{graphicx}
\usepackage{enumitem,stackrel}
\usepackage{color}
\usepackage{mathtools,amssymb,bm}
\usepackage{amstext}
\usepackage{array}
\usepackage{hyperref}
\theoremstyle{remark}

\ifCLASSOPTIONcaptionsoff
  \usepackage[nomarkers]{endfloat}
 \let\MYoriglatexcaption\caption
 \renewcommand{\caption}[2][\relax]{\MYoriglatexcaption[#2]{#2}}
\fi
\newcommand{\RN}[1]{%
	\textup{\uppercase\expandafter{\romannumeral#1}}%
}
\newtheoremstyle{mystyle}%                % Name
  {}%                                     % Space above
  {}%                                     % Space below
  {\itshape}%                                     % Body font
  {}%                                     % Indent amount
  {\bfseries}%                            % Theorem head font
  {.}%                                    % Punctuation after theorem head
  { }%                                    % Space after theorem head, ' ', or \newline
  {}%                                     % Theorem head spec (can be left empty, meaning `normal')
\theoremstyle{mystyle}

{}
{}
{}
{}
{}
{}
\usepackage{epstopdf}
\begin{document}
%
%\onecolumn
% paper title
% can use linebreaks \\ within to get better formatting as desired
\title{Super-Resolution DOA Estimation for Wideband Signals using Arbitrary Linear Arrays  without Focusing Matrices}

\author{Milad~Javadzadeh Jirhandeh, Mohammad Hossein~Kahaei \thanks{The authors are with the School of Electrical Engineering, Iran University of Science \& Technology, Tehran 16846-13114, Iran (e-mail: milad\_javadzade@elec.iust.ac.ir; kahaei@iust.ac.ir).}}%

% make the title area
\maketitle
\begin{abstract}
    We focus on developing an effective Direction Of Arrival (DOA) estimation method for wideband sources based on the gridless sparse concept. Previous coherent methods have been designed by dividing wideband frequencies into a few subbands which are transferred to a reference subband using a focusing matrices. In this work, as opposed to the previous techniques, we propose a convex optimization problem that leads to an accurate wideband DOA estimation method with no need for any focusing matrix. Moreover, in this method no initial DOA estimates are required and it can be used for any arbitrary linear arrays. Numerical simulations show that in comparison to some well-known techniques, the proposed method generates outstanding accuracy and better robustness to noise. The effectiveness of the method is also veriﬁed in presence of close adjacent sources.
\end{abstract}
% Note that keywords are not normally used for peerreview papers.

% For peer review papers, you can put extra information on the cover
% page as needed:
% \ifCLASSOPTIONpeerreview
% \begin{center} \bfseries EDICS Category: 3-BBND \end{center}
% \fi
%
% For peerreview papers, this IEEEtran command inserts a page break and
% creates the second title. It will be ignored for other modes.
\IEEEpeerreviewmaketitle

\section{Introduction}\label{sec:Introduction}
\IEEEPARstart{D}{irection} of arrival estimation is employed in many applications such as seismic \cite{1}, acoustic \cite{2},  sonar \cite{3}, radar \cite{4}, and wireless communications \cite{5}. However, most of the techniques are devoted to narrowband sources, while in many applications the received signals are wideband for which narrowband methods would result in inaccurate and irreliable estimates. This, as a result,
has motivated more research to develop appropriate techniques for wideband signals.  To do so, a conventional approach is to divide a wideband  output signal of an array into some narrowband subbands using a filter bank or by employing the Discrete Fourier Transformation (DFT). Then, to combine the joint information of these narrowbands, two conventional methods are classified into Incoherent Signal Subspace Methods (ISSMs), and Coherent Signal Subspace Methods (CSSMs).
%\IEEEPARstart{W}{ideband } DOA estimation is employed in many applications such as seismic \cite{1}, acoustic \cite{2},  sonar \cite{3}, radar \cite{4}, and wireless communications \cite{5}. However, most of the techniques are devoted to narrowband sources, and MUSIC \cite{6} is one of the most popular techniques. However, in many applications like acoustics signals, seismic waves, and some radars, the received signals are wideband and narrowband methods would be unable to provide a reliable estimation. Therefore, wideband estimation methods have been developed. A conventional approach is to divide the wideband output signal of the array into narrowband components by using filter banks or employing Discrete Fourier Transformation (DFT). Based on the approach to combine the joint information that exists in these narrowband components, the conventional methods are classified into Incoherent Signal Subspace Methods (ISSMs), and Coherent Signal Subspace Methods (CSSMs).

In the ISSM, a narrowband DOA estimator is applied to all narrowband subbands and the final DOA estimate is then calculated by incoherently averaging the respective results \cite{8,9}. These methods show poor performance for low Signal to Noise Ratio (SNR) signals and the DOA estimate  is sensitive to any high-magnitude error that may occur in each subband.
In the CSSM \cite{10,11}, the center frequencies of all narrowband subbands are focused on a reference subband by using a focusing matrix. Then an accurate narrowband DOA estimator such as the MUSIC \cite{6} is applied to the reference frequency for all the focused subbands to obtain the final DOA estimate. The Rotational Signal Subspace (RSS)  is one of the popular CSSMs \cite{11}. However, a major difficulty with the CSSM is how to design the corresponding focusing matrices. Moreover, most of these methods need an initial value for DOA estimation which is considered a practical drawback.
To overcome these limitations, some methods such as the Test of Orthogonality of Projected Subspaces (TOPS) \cite{12}, and Weighted Squared TOPS (WS-TOPS) \cite{7} have been developed in  signal and noise subspaces for different narrowband subbands. Although these methods need no focusing matrix for the estimation procedure, they need a large number of snapshots to offer acceptable performance, which might not be available in some applications or impose a large computational burden.

As another approach, in \cite{13} and \cite{14}, the DOAs are estimated based on Compressive Sensing (CS) theory by employing the joint sparsity in different frequency bins. These methods can lead to high accuracy estimates for a low number of snapshots. To employ the CS concept,  the DOA space is discretized by a grid and the DOA estimates would be located on the nearest corresponding grid point. This, as a result, would limit the resolution of the estimates to the resolution of the grid. It is known that in practice the true DOAs could be located off the grid cells  (off-grid) which could lead to an unavoidable grid mismatch error. To overcome this limitation,  some off-grid methods  estimate the DOAs and the grid simultaneously \cite{15,16}. An off-grid method for wideband DOA estimation has been reported in \cite{17} based on Sparse Bayesian Learning (SBL). These methods normally involve nonconvex optimization problems whose  global convergence can not be guaranteed. Moreover, they need an initial grid.

To overcome the mentioned problems, Candés and Fernandez-Granda \cite{18} extended the discrete sparse approach to a continuous case by introducing the super-resolution concept, which ultimately leads to solving a convex optimization problem. In \cite{19}, the super-resolution notion was introduced as a gridless sparse method for line spectral estimation. The Application of the latter approach was later developed in \cite{20} and \cite{21} for narrowband DOA estimation. It should be noted that all of these methods are performed using regular arrays such as Uniform Linear Arrays (ULA) or Sparse Linear Arrays (SLA). In some applications \cite{22}, however, making use of such arrays is impossible as the sensors should be located in arbitrary positions. The employment of irregular array architectures provided a viable solution in \cite{23}. In \cite{24}, a sparse gridless arbitrary sampling-based method is developed for line spectral estimation based on Prolate Spheroidal Wave Functions (PSWFs). This method uses the Single Measurement Vector (SMV) model and can be applied for narrowband DOA estimation with arbitrary linear arrays. We extended this method for Multiple Measurement Vectors (MMVs) case in \cite{25}.

In this paper, we propose a Super-Resolution Wideband DOA (SRW-DOA) estimator using arbitrary linear arrays which leads to a convex optimization problem. Also,  the resulting optimization problem is solved by Semi-Definite Programming (SDP) developed in our previous work \cite{25}. The major point  is that in this method no focusing matrix and initial estimates  are required.  Our numerical results show that the proposed method offers high accuracy estimates with more robustness to noise compared to the conventional ones. Moreover, this method is useful for DOA estimation of adjacent sources.

The paper is organized as follows. In Section \ref{sec:data model}, a data model is introduced. In Section \ref{sec:proposed1}, the gridless sparse method is proposed for wideband DOA estimation. Numerical results are shown in \ref{sec:simulations}, and Section \ref{sec:coclusion} concludes the paper.

In this work, matrices and vectors are respectively represented by uppercase and lowercase bold letters, ${\left(  \cdot  \right)^T}$ and ${\left(  \cdot  \right)^H}$ denote transpose and conjugate transpose operators, ${\left\| {\cdot} \right\|_F}$ and ${\left\| {\cdot} \right\|_{\cal A}}$ are the Frobenius and atomic norm, respectively, $\mathop {\left\| {\cdot} \right\|}\nolimits_2$ defines  the $\ell _ {2} $ norm of a vector, $diag({\cdot})$ and $toep({\cdot})$ convert a vector to diagonal and Toeplitz matrices, respectively and finally, $conv({\cdot})$, $trace({\cdot})$, and $tan({\cdot})$ represent the convex hull, trace, and tangent operator, respectively.

\section{Data model}\label{sec:data model}
We consider an arbitrary linear array which consists of $M$ omnidirectional sensors located with arbitrary distances. Assuming the first sensor as the reference, the distance of the $m$th sensor from the reference is ${r_m}$, and for the reference sensor, ${r_1 = 0}$. Also, there exist $K$ sources with directions ${\theta _k}, k = 1, \ldots ,K$, emitting different wideband signals. It is also considered that these sources are fixed during the observation time. By applying the DFT to the wideband signal received by each sensor which are sampled by the Nyquist rate, we get $J$ frequency bins. Then, the $j$th frequency bin for $K$ received signals at  the $m$th sensor is presented as
\begin{align}\label{formulation:1}
{X_{m,j}} = \mathop \sum \limits_{k = 1}^K {S_{k,j}}{\rm{exp}}\left( { - \frac{{i2\pi {r_m}\sin \left( {{\theta _k}} \right)}}{{{\gamma _j}}}} \right) + {N_{m,j}},\\
 j = 1, \ldots ,J,\;\;m = 1, \ldots ,M, \nonumber
\end{align}
where ${S_{k,j}}$  shows the emitted signal by the $k$th source, ${\gamma _j}$ is the wavelength of the $j$th frequency, and ${N_{m,j}}$ is the corresponding noise at the $m$th sensor. By defining ${\alpha _j} \buildrel \Delta \over = \frac{2}{{{\gamma _j}}}$, the spatial DOA frequencies ${f_k} = \frac{1}{2}\sin \left( {{\theta _k}} \right) \in \left[ { - \frac{1}{2},\frac{1}{2}} \right]$, and using  ${X_{m,j}}$ and  ${N_{m,j}}$ as the elements of matrices  $\bm{X} \in \mathbb{C}{^{M \times J}}$ and $\bm{N} \in \mathbb{C}{^{M \times J}}$, respectively, (\ref{formulation:1}) is arranged in matrix form as
\begin{equation}\label{formulation:2}
\bm{X} = \mathop \sum \limits_{k = 1}^K {\beta _k}\left[ {{\bm{a}_1}\left( {{f_k}} \right), \ldots ,{\bm{a}_J}\left( {{f_k}} \right)} \right]diag\left( {{\bm{c}_k}} \right) + \bm{N},
\end{equation}
where we assume $\|\bm{N}\|_F^2 \le {\sigma ^2}$,
$$\begin{array}{l}
{\beta _k} = \;\mathop {\left\| {\left[ {\begin{array}{*{20}{c}}
{{s_{k,1}}}\\
 \vdots \\
{{s_{k,J}}}
\end{array}} \right]} \right\|}\nolimits_2 \in \mathbb{R}{^{+}},\;\;{{\bm{c}}_k} =  {\left[ {\begin{array}{*{20}{c}}
{{s_{k,1}}}\\
 \vdots \\
{{s_{k,J}}}
\end{array}} \right]} /{\beta _k} \in \mathbb{C}{^{J \times 1}},\\
\end{array} $$
and the $j$th steering vector  for a general spatial DOA frequency of $f$ is given by
\begin{equation}\label{formulation:3}
{\bm{a}_j}\left( f \right) = \left[ {\begin{array}{*{20}{c}}
{exp\left( { - i2\pi {\alpha _j}{r_1}f\;} \right)}\\
{\begin{array}{*{20}{c}}
{exp\left( { - i2\pi {\alpha _j}{r_2}f\;} \right)}\\
 \vdots
\end{array}}\\
{exp\left( { - i2\pi {\alpha _j}{r_M}f\;} \right)}
\end{array}} \right] \in \mathbb{C}{^{M \times 1}}.
\end{equation}
To present (\ref{formulation:2}) by a model in order to incorporate in the gridless sparse problem, we define a general steering vector using $\bm{a}_j(f)$'s by defining the set,
$${\cal R} = \left\{ {{\alpha _j}{r_m} \in \mathbb{R}{^{+}}\;:\;\;j = 1, \ldots ,J,\;\;m = 1, \ldots ,M} \right\},$$
and the vector ${\left[ {{{\tilde r}_1}, \ldots ,\;{{\tilde r}_{\tilde M}}} \right]^T}$,
 whose elements
$ {\tilde r_{\tilde m}}={\alpha _j}{r_m}, \tilde m = 1, \ldots ,\;\tilde M$ ($M \le \tilde M \le J \times M$) are the sorted members of  ${\cal R}$ in ascending order after deleting the recurring elements. Then, the general steering vector is defined as
\begin{equation}\label{formulation:4}
\bm{a}\left( f \right) = \left[ {\begin{array}{*{20}{c}}
{exp\left( { - i2\pi {{\tilde r}_1}f\;} \right)}\\
{\begin{array}{*{20}{c}}
{exp\left( { - i2\pi {{\tilde r}_2}f\;} \right)}\\
 \vdots
\end{array}}\\
{exp\left( { - i2\pi {{\tilde r}_{\tilde M}}f\;} \right)}
\end{array}} \right] \in {\mathbb{C}^{\tilde M \times 1}},
\end{equation}
which corresponds to the steering vector of an arbitrary linear array with $\tilde M$  sensors spaced with the distances ${\tilde r_{\tilde m}}$ from each other. It can be seen that each steering vector $\bm{a}_j(f)$ is a sub-vector with $M$ elements of $\bm{a}(f)$.

In order to represent (\ref{formulation:2}) based on $\bm{a}(f)$, we define the map $\chi:{\mathbb{C}^{\tilde M \times J}} \to {\mathbb{C}^{M \times J}}$, which relates the creating matrix  $\left[ {{\bm{a}_1}\left( f \right), \ldots ,{\bm{a}_J}\left( f \right)} \right] \in {\mathbb{C}^{M \times J}}$ in (\ref{formulation:2}) to $\bm{a}(f)$ in (\ref{formulation:4}). Then, for any matrix $\bm{Z} \in {{\mathbb{C}^{\tilde M \times J}}}$, this map yields,
%\begin{equation}\label{formulation:5}
%\chi {\left( \bm{Z} \right)_{mj}}??????OR \chi {\left( \bm{Z}_{m,j}\right)} = {\bm{Z}_{\tilde m,j}},\;\;\;
%\end{equation}
\begin{equation}\label{formulation:5}
(\chi {\left( \bm{Z} \right))_{m,j}} = {\bm{Z}_{\tilde m,j}},\;\;\;for, \;\;\;{\alpha _j}{r_m} = {\tilde r_{\tilde m}}.
\end{equation}
 Thus, we get
 \begin{equation}\label{formulation:6000}
\left[ {{\bm{a}_1}\left( f \right), \ldots ,{\bm{a}_J}\left( f \right)} \right] = \chi \left( {\bm{a}\left( f \right){{\left( {{\bm{1}_{J \times 1}}} \right)}^T}} \right),
\end{equation}
where ${\bm{1}_{J \times 1}}$ is an all-one vector. Now, using $\chi$ in (\ref{formulation:5}) and (\ref{formulation:6000}), we can represent (\ref{formulation:2}) as
\begin{align}\label{formulation:6}
& \bm{X} = \mathop \sum \limits_{k = 1}^K {\beta _k}\chi \left( {\bm{a}\left( {{f_k}} \right){{\left( {{\bm{1}_{J \times 1}}} \right)}^T}} \right)diag\left( {{\bm{c}_k}} \right) + \bm{N}\\
 & = \chi \left( {\mathop \sum \limits_{k = 1}^K {\beta _k}\bm{a}\left( {{f_k}} \right){{\left( {{\bm{1}_{J \times 1}}} \right)}^T}diag\left( {{\bm{c}_k}} \right)} \right) +\bm{N}\nonumber\\
 & = \chi \left( {\mathop \sum \limits_{k = 1}^K {\beta _k}\bm{a}\left( {{f_k}} \right)\bm{c}_k^T} \right) + \bm{N}.\nonumber
\end{align}
In other words, we can present the model,
\begin{equation}\label{formulation:7}
\bm{X} = \chi \left( \bm{Z} \right) + \bm{N},\;\;\; \bm{Z} = \mathop \sum \limits_{k = 1}^K {\beta _k}\bm{a}\left( {{f_k}} \right)\bm{c}_k^T,
\end{equation}
to propose a gridless sparse method for wideband DOA estimation.
\section{Super-Resolution Wideband DOA Estimator}\label{sec:proposed1}
In the following, we elaborate on the proposed SRW-DOA estimator. According to the structure of $\bm{Z}$ in (\ref{formulation:7}), we state that the building blocks of this matrix are the members of the following set of atoms,
\begin{equation}
\begin{array}{l}
{\cal A} = \{{\bm{A}}\left( {f,{\bm{c}}} \right) =  {{\bm{a}}\left( f \right)}{\bm{c}^T}\in{\mathbb{C}^{\tilde M \times J}}\;\; {\rm{|}}\;\\
\;\;\;\;\;\;\;\;\;\;f \in \left[ { - \frac{1}{2},\frac{1}{2}} \right],{\bm{c}} \in\mathbb{C}^{J\times 1} ,\;{\left\| {\bm{c}} \right\|_2} = 1\}\nonumber.
\end{array}
\end{equation}
The atomic norm of $\bm{Z}$ is the possible minimum number of the atoms of ${\cal A}$, by which $\bm{Z}$ can be constructed. The atomic norm is defined according to \cite{21} and \cite{26} as
\begin{align}\label{formulation:8}
&{\|\bm{Z}\|_{\cal A}} = {\rm{inf}}\{ t > 0:\bm{Z} \in t\;conv\left( {\cal A} \right)\} \\
& = \mathop {\inf }\limits_{{f_k},{\beta _k},{\bm{c}_k}} \left\{ {\mathop \sum \nolimits_k {\beta _k}\;:\bm{Z} = \mathop \sum \nolimits_k {\beta _k}\bm{A}\left( {{f_k},{\bm{c}_k}} \right),\;\;{\beta _k} > 0\;} \right\}\nonumber.
\end{align}
Since $K$ is finite, the number of constructing atoms of $\bm{Z}$ will be finite, as well. As a result, to recover $\bm{Z}$, we propose the following Atomic Norm Minimization (ANM) problem,
\begin{align}\label{formulation:9}
&\mathop {\min }\limits_{\bm{Z} \in {^{\tilde M \times J}}} \;{\|\bm{Z}\|_{\cal A}}\\
&s.t. \;\;\; {\|\bm{X} - \chi\left( \bm{Z} \right)\|}_F \le \sigma \nonumber.
\end{align}
Noting that $\bm{a}\left( f \right)$ shows the steering vector of an arbitrary linear array with $\tilde M$  sensors with arbitrary distances ${\tilde r_{\tilde m}}$ , (\ref{formulation:9}) can be considered an ANM problem with arbitrary sampling, which has been solved in \cite{25} by   using PSWFs as an SDP.
Considering ${\mathbb{L}^2}$ as the set of all square-integrable functions on $\left[ { - \frac{1}{2},\frac{1}{2}} \right]$  and $c = \pi {\tilde r_{\tilde M}}$, for any  $r \in {\mathbb{L}^2}$, PSWFs are defined as the eigenfunctions of the linear map $\xi :{\mathbb{L}^2} \to {\mathbb{L}^2}$ as
\begin{equation}\label{formulation:10}
  (\xi r)(\tau)=\int^1_{-1}e^{ic\zeta \tau}r(\zeta)d\zeta, \:\:\:\:\:\:\:\forall \tau\in [-1,1].
\end{equation}
Therefore, the PSWF, ${\varphi _l}$, should satisfy $\xi {\varphi _l} = {\lambda _l}{\varphi _l}$, where ${\lambda _l}$ is the $l$th eigenvalue of $\xi$. According to \cite{1000}, as the amplitude of the eigenvalues larger than $\frac{{2c}}{\pi }$ tends to zero,  we can limit the number of required PSWFs to $d = \left\{ {min\frac{l}{2}\;|\;\left| {{\lambda _l}} \right| <  \epsilon} \right\}$, where $\epsilon$ defines the desired precision \cite{27}. Then, using the latter PSWFs, we can represent (\ref{formulation:9}) as an SDP \cite{25} by
\begin{align}\label{formulation:10}
  \underset{\underset{v_1,\dots,v_d\in \mathbb{C},v_0\in\mathbb{R}}{\boldsymbol W\in \mathbb{R}^{J\times J},\bm{Z} \in \mathbb{C}^{\tilde M\times J}}} {min} & \:\:\:\:\frac{1}{2}(trace(\boldsymbol W)+ \bm{Q}_{1,1})  \\
 s.t. \:\:\:\:\:\:\:\:& \left[
           \begin{array}{cc}
             \boldsymbol W & \boldsymbol Z^H \\
             \boldsymbol Z & \boldsymbol Q \\
           \end{array}
         \right]\succeq0,\nonumber\\
        & {\|\bm{X} - \chi\left( \bm{Z} \right)\|}_F \le \sigma,   \nonumber\\
 &  \boldsymbol Q_{q,p}=\boldsymbol h_{qp}^T\boldsymbol{\Phi}^{-1}[v_d^H,\dots,v_1^H,v_0,v_1,\dots,v_d]^T, \nonumber\\
  & \boldsymbol T:= toep([v_0,v_1,\dots,v_d]), \nonumber\\
    &\Psi(\boldsymbol T):=\tan^2(\frac{{c}}{2 d})(\boldsymbol J_1+\boldsymbol J_2)\boldsymbol T(\boldsymbol J_1+\boldsymbol J_2)^H \nonumber \\
     & -(\boldsymbol J_1-\boldsymbol J_2)\boldsymbol T(\boldsymbol J_1-\boldsymbol J_2)^H\succeq0,\nonumber
\end{align}
where ${\bf{\Phi }} \in {\mathbb{R}^{\left( {2d + 1} \right) \times \left( {2d + 1} \right)}}$ and $\bm{h}_{qp} \in {\mathbb{R}^{\left( {2d + 1} \right)}}$  are defined as
\begin{align}
& {{\bf{\Phi }}_{q,l}} = {\varphi _{l - 1}}\left( {\frac{{q - d - 1}}{d}} \right), \nonumber \\
& {\bm{h}_{qp}}\left( l \right) = {\varphi _{l - 1}}\left( {\frac{{{{\tilde r}_q} - {{\tilde r}_p}}}{{{{\tilde r}_{\tilde M}}}}} \right),\nonumber
\end{align}
and ${\bm{J}_1} = \left[ {{\bm{I}_d},{\bm{0}_{d \times 1}}} \right]$, ${\bm{J}_2} = \left[ {{\bm{0}_{d \times 1}},{\bm{I}_d}} \right]$, ${\bm{I}_d}$ is a $d \times d$ identity matrix, and ${\bm{0}_{d \times 1}}$  is an all-zero vector.
By solving the SDP problem in (\ref{formulation:10}), the Toeplitz response matrix $\bm{T}$ is obtained which is singular with rank $K$  \cite{25}. Therefore, using the {Proney'}s method \cite{28}, we can recover the spatial DOA frequencies  ${f_k}'$s and accordingly the DOAs ${\theta _k}'$s.

\section{Numerical simulations}\label{sec:simulations}
We compare the performance of the proposed SRW-DOA estimator with other existing wideband DOA estimation methods including RSS \cite{11}, WS-TOPS \cite{7}, and SBL \cite{17} in different experiments. We consider an underwater environment with a wave propagation speed of $1500 m/s$. Each source emits random wideband signals that have a bandwidth of $167 Hz$ and a center frequency of $500 Hz$. The received signal in each sensor is sampled at the rate of $2000 Hz$ with $512$ samples. An arbitrary linear array consisting of $M=8$ sensors is considered, which ${r_m}'$s are drawn randomly from a uniform distribution on the interval $\left( {0,M\frac{{{\gamma _c}}}{2}} \right)$.
The distances of the first and last sensors from the reference sensor are ${r_0} = 0$ and ${r_M} = M\frac{{{\gamma _c}}}{2}$, respectively, and ${\gamma _c}$ shows the wavelength at the center frequency.
The measuring noise in (\ref{formulation:1}) is zero-mean Gaussian with variance ${\sigma ^2}$  and the SNR is defined as $\frac{{\|\bm{X} - \bm{N}\|_F^2}}{{{\sigma ^2}}}$. The initial DOA estimates for the RSS are selected as the true DOAs added up with a uniformly distributed random error within $ \pm {2^\circ }$. The initial grid for the SBL is defined as a ${1^\circ }$ discretized uniform grid. The number of sources $K$ is assumed to be known. Accordingly, in the RSS and WS-TOPS, the $K$ largest peaks in the spatial frequency are selected as the DOA estimates, while in the SBL and SRW-DOA, the $K$ largest components are taken as the DOA's. Simulation results are presented by averaging $100$ independent trials of each experiment. The Root Mean-Square Error (RMSE) is defined as  {\tiny $ RMSE = \frac{1}{{100}}\mathop \sum \limits_{i = 1}^{100} \left({\frac{1}{K}\mathop \sum \limits_{k = 1}^K {{\left( {{\theta _k} - {{\hat \theta }_{k,i}}} \right)}^2}}\right)^{1/2}$}, where ${\hat \theta _{k,i}}$ is the estimated DOA for the $k_{th}$ source at the $i_{th}$ trial.

In the first experiment, we consider $K=3$ sources with DOAs ${\theta _1} =  - {5^\circ }$, ${\theta _2} = {15^\circ }$, and ${\theta _3} = {40^\circ }$. The number of frequency bins is $J=10$. The error margin is accepted for each DOA estimate if the difference between each estimate and its true value is less than ${5^\circ }$.

\begin{figure}[!t]
\centering
\includegraphics[width=75mm]{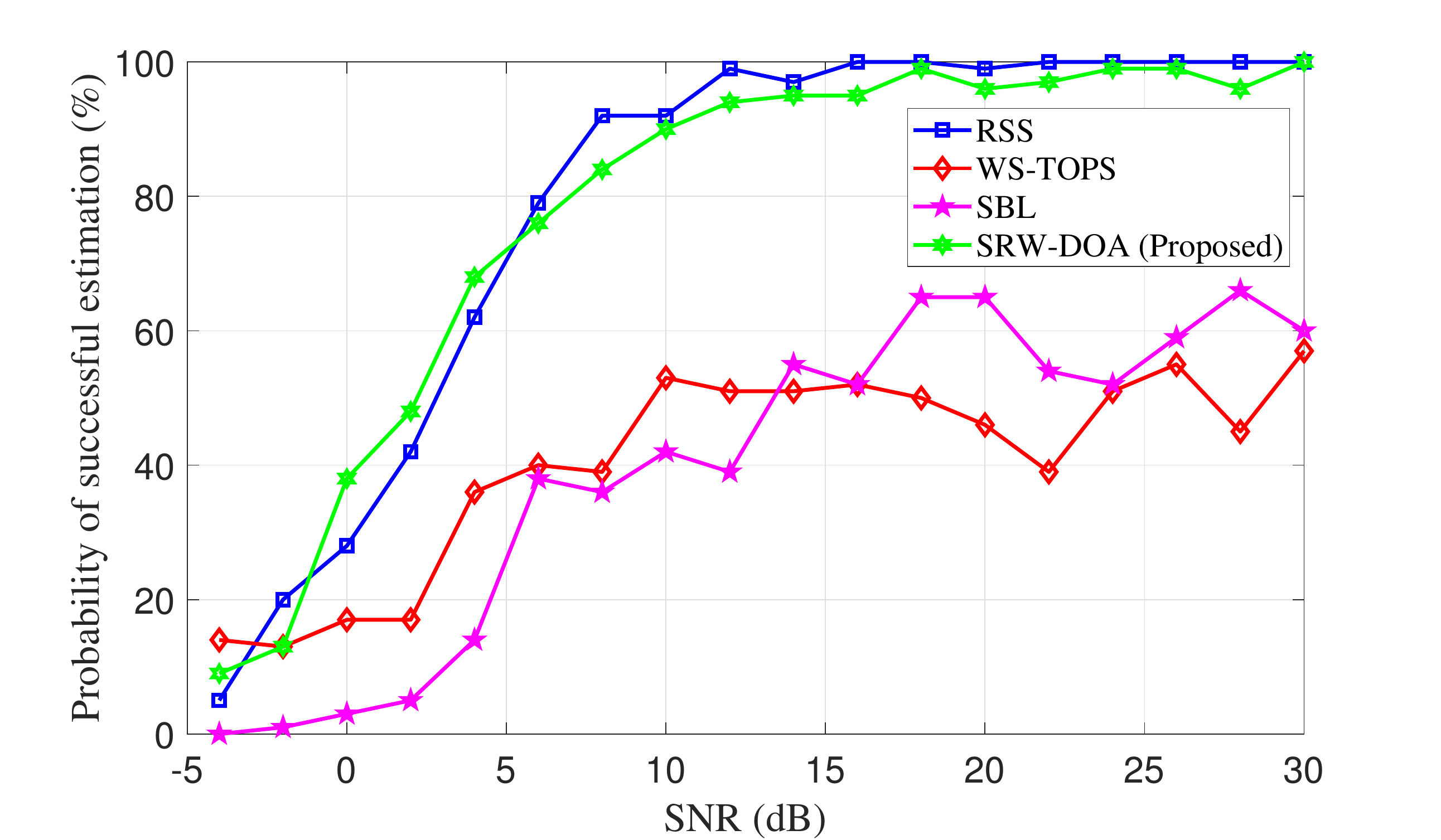}
\caption{Probability of successful estimation v.s. SNR for K=3, ${\theta _1} =  - {5^\circ }$, ${\theta _2} = {15^\circ }$, and ${\theta _3} = {40^\circ }$.}\label{figure1}
\end{figure}

\begin{figure}[!t]
\centering
\includegraphics[width=75mm]{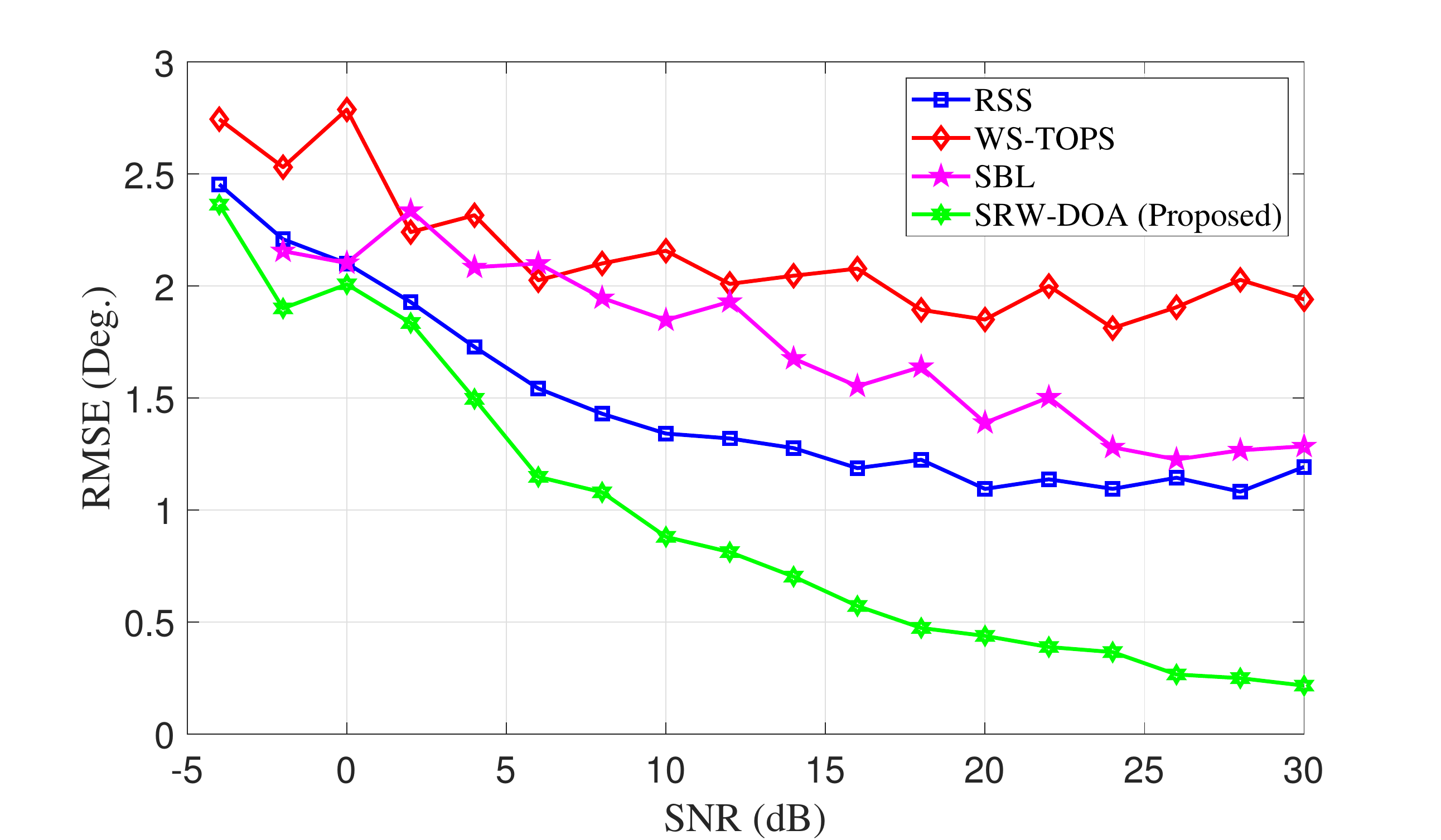}
\caption{RMSEs v.s. SNR for K=3, ${\theta _1} =  - {5^\circ }$, ${\theta _2} = {15^\circ }$, and ${\theta _3} = {40^\circ }$.}\label{figure2}
\end{figure}

The probability of successful estimation is shown in Fig. \ref{figure1} for different SNRs. As seen, both RSS and  SRW-DOA considerably outperform the WS-TOPS and SBL.
 Also, the respective RMSEs in Fig. \ref{figure2} show that the SRW-DOA generates lower errors.

\begin{figure}[!t]
\centering
\includegraphics[width=75mm]{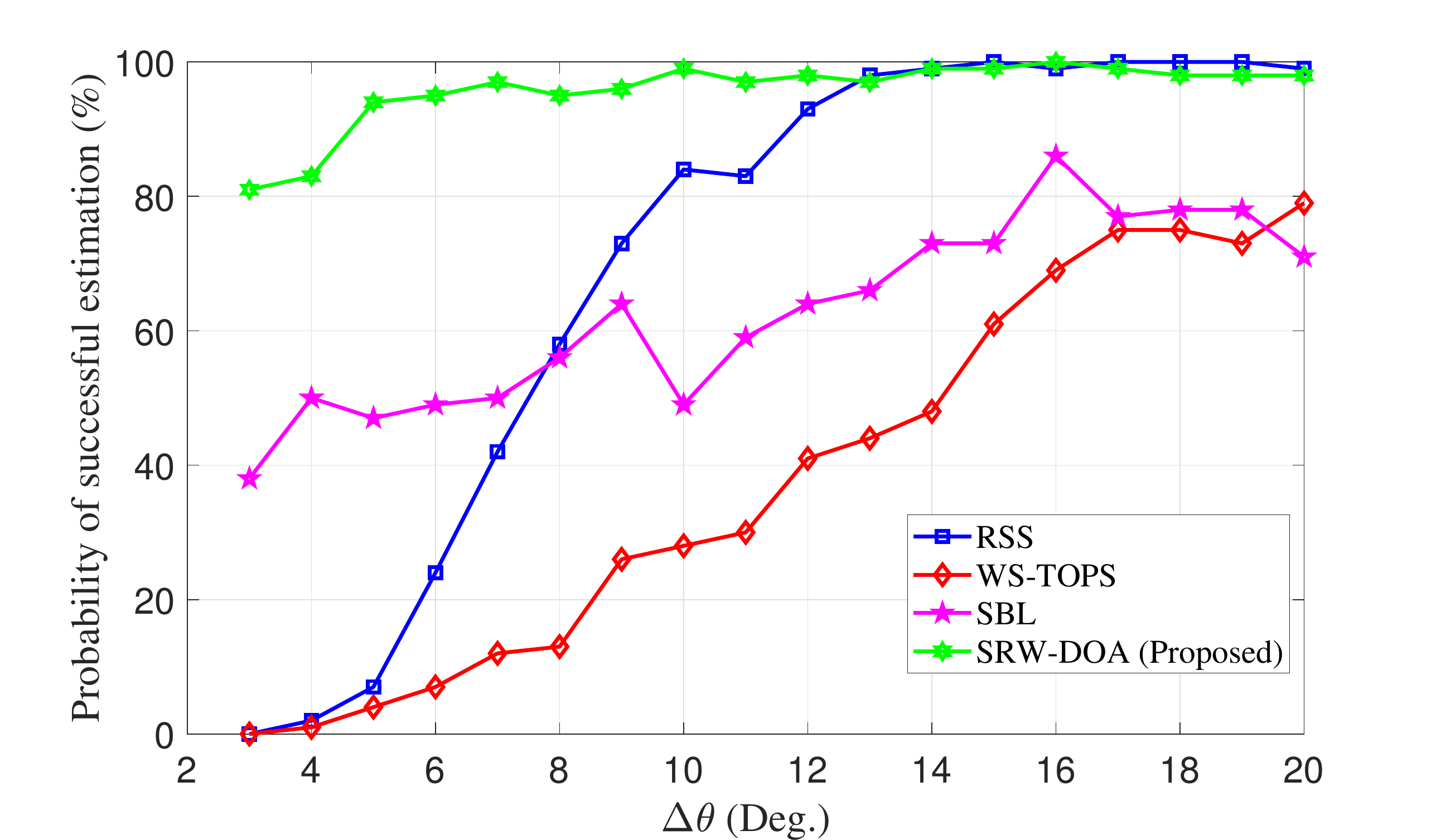}
\caption{Probability of successful estimation v.s. $\Delta\theta$ at $SNR=10 dB$.}\label{figure3}
\end{figure}

\begin{figure}[!t]
\centering
\includegraphics[width=75mm]{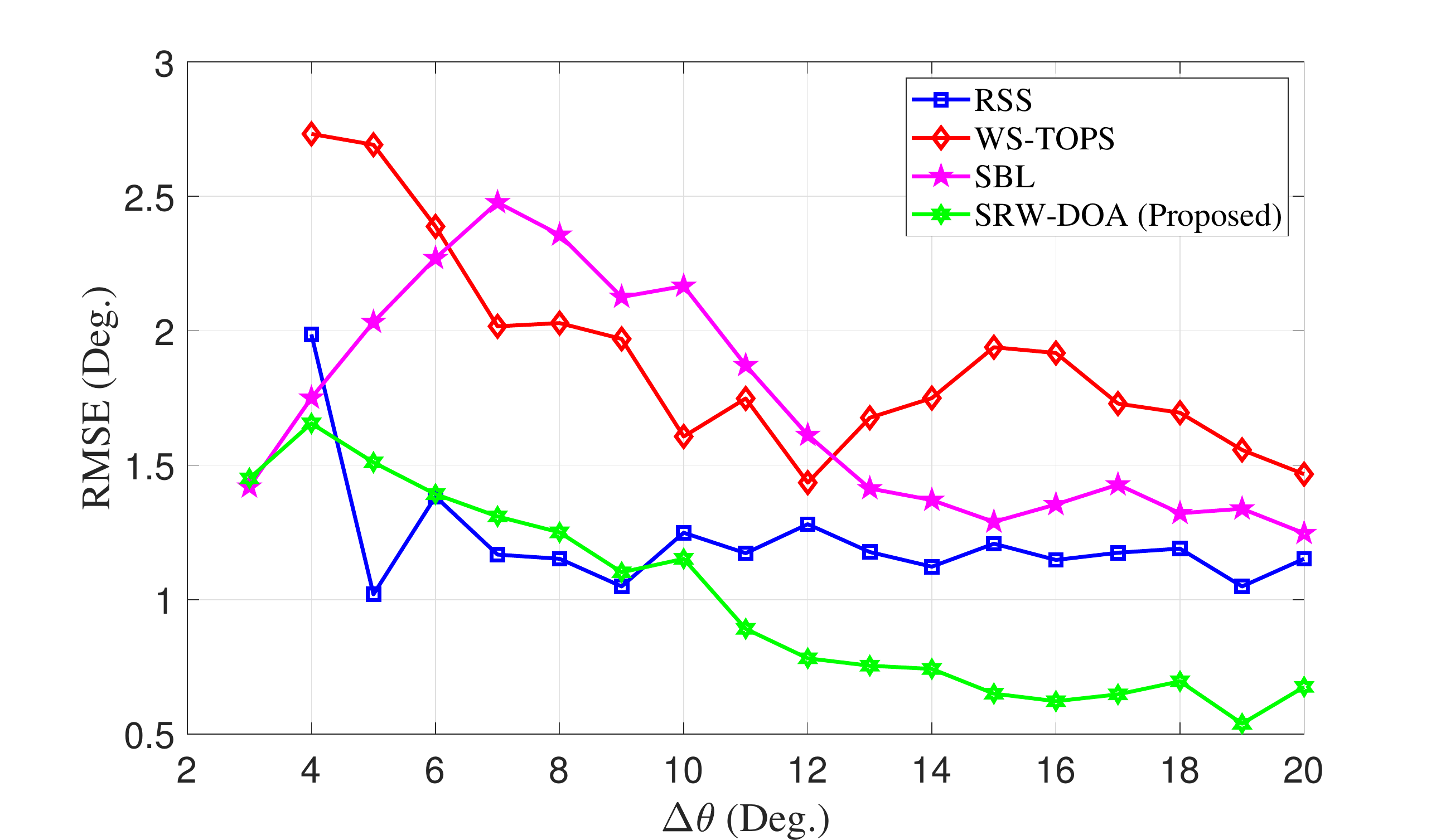}
\caption{RMSEs v.s. $\Delta\theta$ at $SNR=10 dB$.}\label{figure4}
\end{figure}

In the second experiment, we study the ability to distinguish two near DOAs. For this purpose, two sources are considered at ${\theta _1} = {10^\circ }$ and ${\theta _2} = {10^\circ } + \Delta\theta$, where $\Delta\theta$ increases from  ${3^\circ }$ to ${20^\circ }$ and $SNR=10 dB$. The probability of successful estimation and RMSEs are presented in Figs. \ref{figure3} and \ref{figure4}, respectively.
From both figures, one can see that the SRW-DOA (Proposed) significantly outperforms the others, even for small values of $\Delta\theta$.

\begin{figure}[!t]
\centering
\includegraphics[width=75mm]{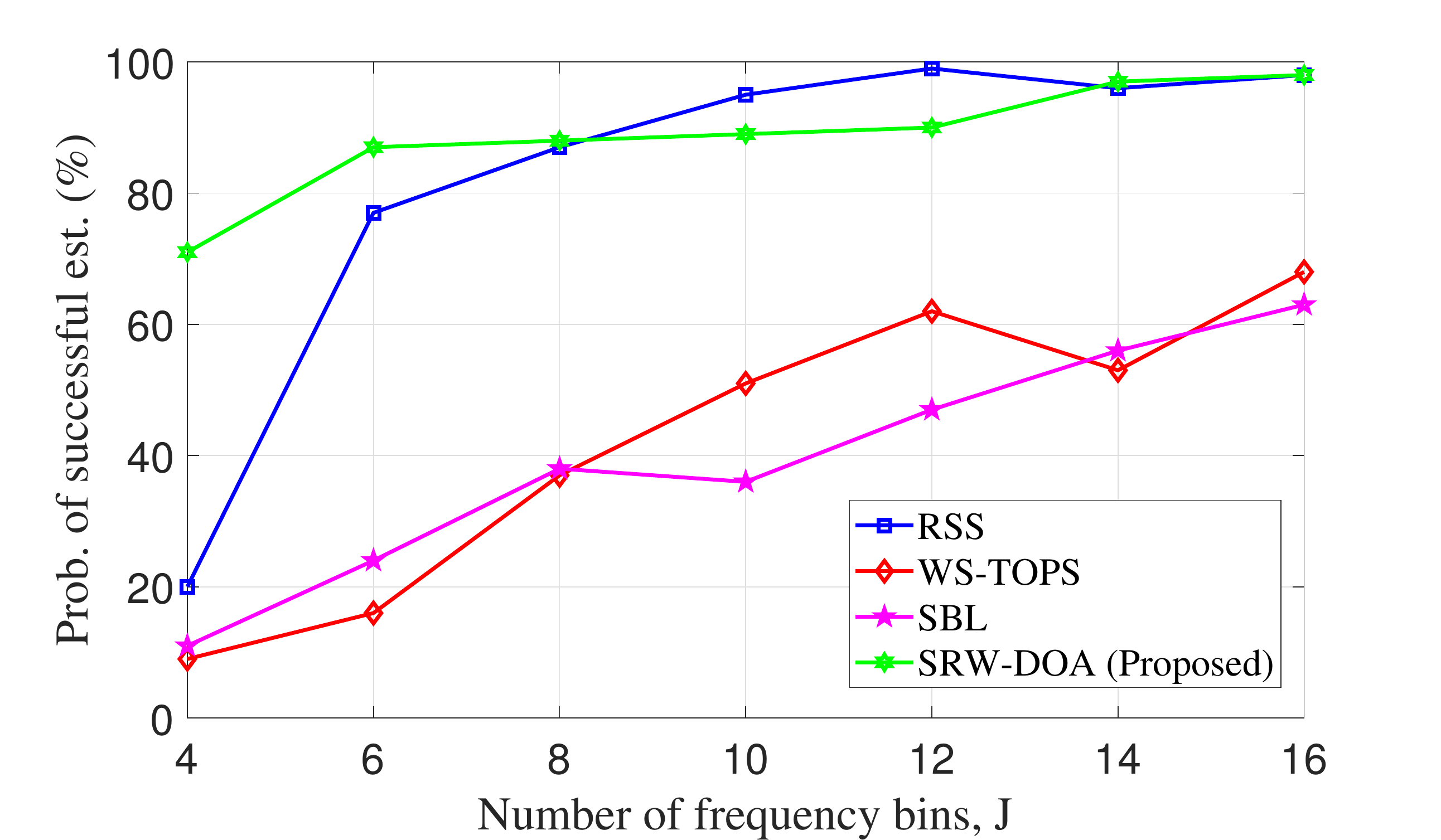}
\caption{Probability of successful estimation v.s. $J$ for K=3, $SNR=10 dB$, and ${\theta _1} =  - {5^\circ }$, ${\theta _2} = {15^\circ }$, and ${\theta _3} = {40^\circ }$.} \label{figure5}
\end{figure}

\begin{figure}[!t]
\centering
\includegraphics[width=75mm]{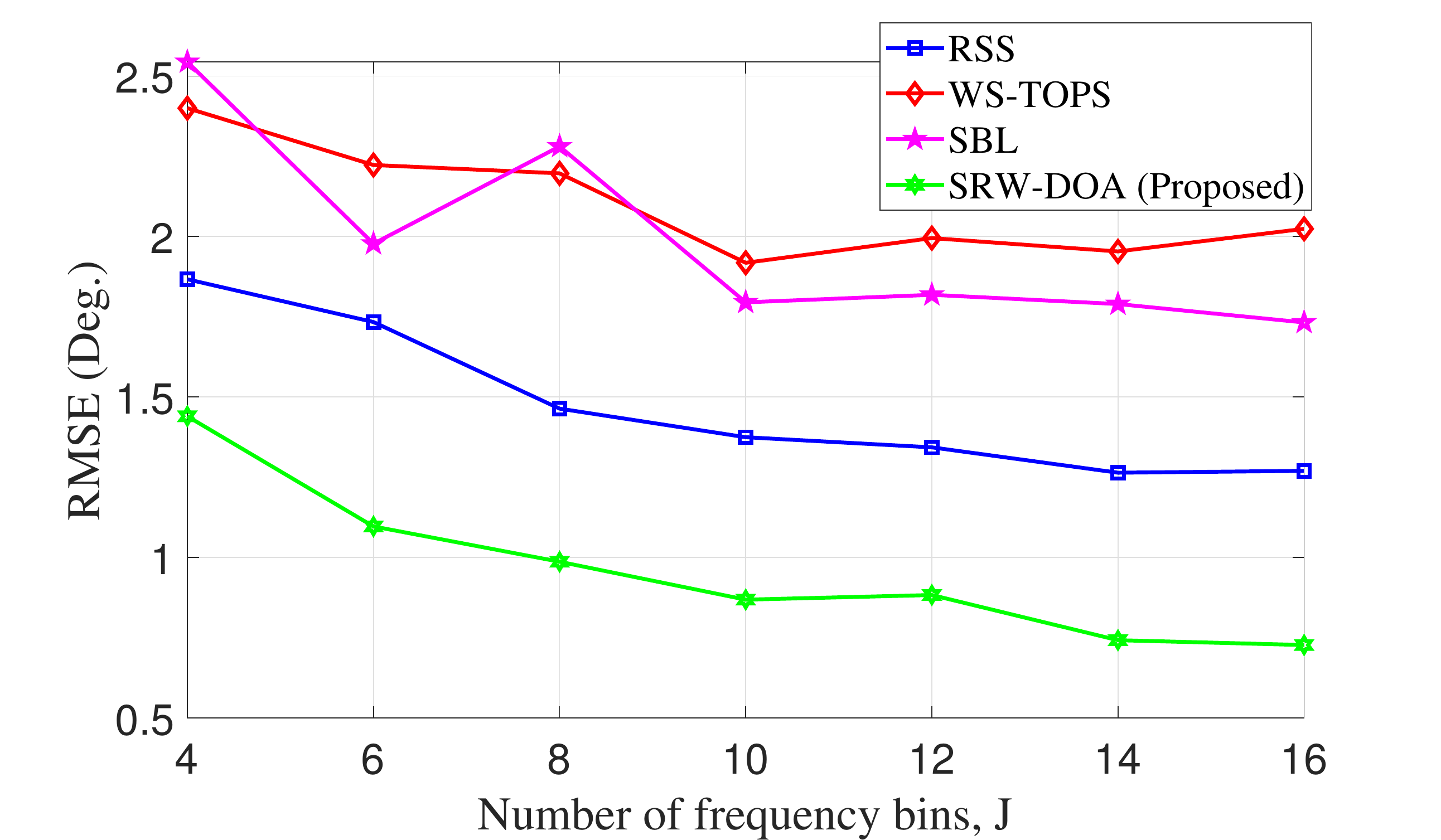}
\caption{RMSEs v.s. $J$ for K=3, $SNR=10 dB$, and ${\theta _1} =  - {5^\circ }$, ${\theta _2} = {15^\circ }$, and ${\theta _3} = {40^\circ }$.}\label{figure6}
\end{figure}

In the last experiment,  the effect of the number of frequency bins on DOA estimates is investigated for $K=3$ sources at ${\theta _1} =  - {5^\circ }$, ${\theta _2} = {15^\circ }$, and ${\theta _3} = {40^\circ }$ and $SNR=10 dB$ for different numbers of frequency bins $J$. The results for different numbers of frequency bins $J$ are shown in Figs. \ref{figure5} and \ref{figure6}, respectively.
From Fig. \ref{figure5}, by increasing the number of frequency bins, the probability of successful estimation increases. Moreover, the RSS and the SRW-DOA offer considerably higher success probabilities compared to WS-TOPS and SBL. Also, Fig. \ref{figure6} shows that the SRW-DOA generates the lowest RMSE for all $J$’s.

\section{Conclusion}\label{sec:coclusion}
We proposed a gridless sparse method for wideband DOA estimation with no need for a focusing matrix or initial estimates. This method can be applied to all arbitrary linear arrays and the corresponding optimization problem is convex. In comparison to RSS, WS-TOPS, and SBL algorithms, the proposed SRW-DOA method showed outstanding performance by generating lower RMSEs, higher probability of successful estimation,  more accurate DOA estimates, and more robustness to noise. Moreover, it showed remarkable effectiveness for estimation of the DOAs of adjacent sources.

% \newpage
\bibliographystyle{ieeetr}
\bibliography{Milad_Main_References}
\balance
%\bibliography{refrence}

\end{document}